\documentclass[prd,reprint,onecolumn,superscriptaddress,amsfonts,nofootinbib]{revtex4-2}
\usepackage{amsmath,amssymb}

\usepackage{slashed}
\usepackage{graphicx}
\usepackage{epsfig}
\usepackage{epstopdf}
\usepackage{adjustbox}

\newcommand{\dd}{{\mathrm{d}}}

\usepackage{xcolor} \definecolor{darkgreen}{rgb}{0,.5,0}
\usepackage[colorlinks,filecolor=blue,citecolor=darkgreen,unicode]{hyperref}

\begin{document}
\title{Target space diffeomorphisms in Poisson sigma models and asymptotic symmetries in 2D dilaton gravities}
\author{Carlos Valc\'arcel}\email{valcarcel.flores@gmail.com}
\affiliation{Instituto de F\'isica - Universidade Federal da Bahia, C\^ampus Universit\'ario de Ondina, 40210-340, Salvador, B.A. Brazil}
\author{Dmitri Vassilevich}\email{dvassil@gmail.com}
\affiliation{CMCC-Universidade Federal do ABC, Av. dos Estados 5001, 09210-580, Santo Andr\'e, S.P., Brazil}
\affiliation{Physics Department, Tomsk State University, 34050, Tomsk, Russia}

\begin{abstract}
The dilaton gravity models in two dimensions, including the Jackiw--Teitelboim model and its deformations, are particular cases of Poisson sigma models. Target space diffeomorphisms map one Poisson sigma model to another. We propose to use these diffeomorphisms to identify asymptotic conditions, boundary actions, and asymptotic symmetries in distinct dilaton gravity models. As an example, we use the asymptotic conditions in Jackiw--Teitelboim gravity to construct an asymptotic problem with Virasoro symmetry in a class of asymptotically Rindler models. We show, that the method can be applied to a wide class of pairs of dilaton gravities and discuss possible generalizations.
\end{abstract}
\maketitle

\section{Introduction}\label{sec:int}
The discovery \cite{Maldacena:2016hyu,Maldacena:2016upp,Jensen:2016pah} that the Sachdev-Ye-Kitaev (SYK) model \cite{Sachdev:1992fk,Kitaev} is a holographic dual of the Jackiw--Teitelboim (JT) gravity \cite{Jackiw:1984,Teitelboim:1983ux} has opened a whole new area of research in two-dimensional holography. The JT gravity is a member of a large family of two-dimensional dilaton gravities \cite{Grumiller:2002nm}. All 2D dilaton gravities are classically and locally quantum integrable \cite{Katanaev:1996bh,Kummer:1996hy}. Quite naturally, many works were dedicated to the search of extensions of the JT/SYK paradigm. In particular, matrix model boundary duals were suggested in \cite{Saad:2019lba} and studied in \cite{Witten:2020wvy,Momeni:2020tyt,Johnson:2020lns,Turiaci:2020fjj,Alishahiha:2020jko}. Generalized sets of asymptotic conditions leading to larger asymptotic symmetry algebras were suggested in \cite{Grumiller:2017qao,Godet:2020xpk}. Some limits of JT gravity were analysed in \cite{Grumiller:2020elf,Gomis:2020wxp}. The dilaton gravity models other than JT were considered from the early days of AdS$_2$/CFT$_1$ correspondence \cite{Cadoni:2001ew}. Some other non-JT holographic correspondences in two dimensions have been considered in \cite{Blake:2016jnn,Cvetic:2016eiv,Hong:2019tsx,Narayan:2020pyj,Cao:2021upq,Afshar:2019axx,Ecker:2021guy,Afshar:2021qvi}. We refer to \cite{Trunin:2020vwy} for an overview of 2D holography. Despite the growing number of examples it is still unclear whether a given dilaton gravity model can possess a consistent set of asymptotic conditions with an interesting symmetry algebra.

The dilaton gravities in two dimensions are particular cases of Poisson sigma models (PSMs) \cite{Schaller:1994es}. For a generic PSM with a specific set of boundary conditions the boundary correlation functions may be expressed through star-products of a quantum mechanics defined on the boundary \cite{Cattaneo:1999fm,Vassilevich:2013ai}, which can already be considered as a sort of a holographic correspondence. The target spaces of PSMs are Poisson manifolds. A target space diffeomorphism maps one PSM to another. We propose to use these diffeomorphisms to identify the models asymptotically rather than in the whole space. If one model has a boundary action and a set of asymptotic conditions leading to a consistent variational problem, the same will apply to the model obtained through a target space diffeomorphism. 

More precisely, our proposal is as follows
A dilaton gravity has an asymptotic region for larger values of the dilaton field $X$ (at $X\to\infty$) if the symplectic leaves of corresponding Poisson structure satisfy some restrictions at this region. These restrictions imply existence of a selected coordinate system on the target space manifold. If another model also satisfy the same restrictions, the coordinate systems may be identified. In other words, these two models may be related through a target space diffeomorphism. Note, that the field $X$ may be used as a coordinate in classical solutions of dilaton gravities. Thus $X\to\infty$ defines an asymptotic region on the base manifold as well. If the diffeomorphism is applied to the asymptotic conditions and to the action (including all boundary terms) which define a consistent variational problem in one model, one automatically gets a consistent variational principle in the second model as well. The asymptotic symmetries also follow this scheme, but there is subtlety. The PSMs are formulated by using the coordinates on target Poisson manifold as field variables. The coordinates do not behave covariantly under diffeomorphisms. As a consequence, the gauge transformation rules are also not covariant, except for the case when all fields are on-shell (see \cite{Bojowald:2004wu} for a discussion of covariance properties of PSMs). Therefore, the correspondence between asymptotic symmetries can be established on-shell only. Fortunately, this restriction does not reduce the asymptotic symmetry algebra.

Dilaton-dependent conformal transformations of the metric, which are particular cases of PSM target space diffeomorphisms, have been used  to construct asymptotic conditions in dilaton gravities in \cite{Ecker:2021guy}.

We illustrate our method with an example of asymptotic correspondence between JT and so-called asymptotically Rindler models. We show, how a known asymptotic problem for JT model gives rise to a new set of consistent asymptotic conditions in asymptotically Rindler models with a Virasoro symmetry algebra. Note, that such asymptotic symmetry is rather non-typical for asymptotically flat space holography.

This paper is organized as follows. In the next section, we review the dilaton gravities in first and second order formulations and their relations to PSMs. The general scheme for establishing asymptotic correspondence between dilaton gravity models is presented in section \ref{sec:asy} where we derive the transformation rules for all fields and parameters of the asymptotic symmetry algebra. Is is worth noting already here that the asymptotic Killing vectors in asymptotically equivalent models coincide. In section \ref{sec:Ex} we derive asymptotic correspondence between JT and Rindler models.  In section \ref{sec:oth} we analyse the presence of asymptotic regions in a two-parameter family of dilaton gravity models. We show, that such regions are present in a large subset of this family including most of physically interesting one. Thus, the proposed method is very general. Section \ref{sec:con} contains some concluding remarks. 

\section{Dilaton gravities and Poisson sigma models}\label{sec:dil}
Practically all\footnote{An even more general class of models was considered in \cite{Grumiller:2002md}. Holographic aspects of these models were studied recently in \cite{Grumiller:2021cwg}.} pure dilaton gravity models in two dimensions are described by the following second order action
\begin{equation}\label{D01}
I_{\mathrm{2nd}} = -\frac{1}{2} \int_{\mathcal M}\mathrm{d}^{2}x\:\sqrt{g}\left[X R-U\left( X\right)\left(\partial X\right)^{2}-2V\left(X\right)\right].
\end{equation}
Here $U$ and $V$ are two arbitrary functions (potentials) of the dilaton field $X$, $R$ is the curvature scalar of two-dimensional metric $g_{\mu\nu}$ on the manifold $\mathcal{M}$. 
In this paper, we work with Euclidean signature models. Modifications to the case of Minkowski signature are straightforward.

The action (\ref{D01}) is classically equivalent to the first-order action
\begin{equation}\label{D02}
I_{\mathrm{1st}}=\int_{\mathcal M}\left[e_{\alpha}\wedge\mathrm{d}X^{\alpha}+\omega\wedge \mathrm{d}X+\epsilon^{\alpha\beta}X_{\alpha}\omega\wedge e_{\beta}-\tfrac{1}{2}\epsilon^{\alpha\beta}\mathcal{V} e_{\alpha}\wedge e_{\beta}\right],
\end{equation}
where $e_\alpha$ and $\omega$ are the zweibein and connection one-forms, respectively. The indices $\alpha,\beta\in \{1,2\}$ are lowered and raised with the Kronecker symbol $\delta_{\alpha\beta}$ while $\epsilon^{\alpha\beta}$ is the antisymmetric Levi-Civita symbol, $\epsilon^{12}=1$. $X^\alpha$ denotes auxiliary fields generating the torsion constrains, and
\begin{equation}
\mathcal{V}=\tfrac 12 U(X) X^\gamma X_\gamma + V(X).\label{calV}
\end{equation}

We like to embed the dilaton gravities in an even more general class of 2-dimensional models. To this end,
let us consider a Poisson manifold\footnote{The interested reader may consult the textbook \cite{Crainic:2021} for an introduction to the Poisson geometry.} $\mathcal{P}$ with local coordinates $X^I$ and a Poisson bi-vector $\Pi^{IJ}(X)$ satisfying the Jacobi identity
\begin{equation}
\Pi^{IL}\partial_L\Pi^{JK}+\Pi^{KL}\partial_L\Pi^{IJ}+\Pi^{JL}\partial_L\Pi^{KI}=0.\label{Jac}
\end{equation}
This bi-vector defines a Poisson bracket
\begin{equation}
\{ F, G\} :=\Pi^{IJ}\partial_I F\, \partial_J G \label{Pbra}
\end{equation}
for $F,G\in C^\infty(\mathcal{P})$ and makes $C^\infty(\mathcal{P})$ a Poisson algebra.

Let us make $\mathcal{P}$ a target space of a sigma model. Then $X^I$ become functions of $x$. I.e., they are interpreted as maps $\mathcal{M}\to\mathcal{P}$. We also take fields $A_I(x)$ which are one-forms on $\mathcal{M}$ with values in the cotangent space of $\mathcal{P}$. The action of a PSM reads \cite{Schaller:1994es}
\begin{equation}
I_{\mathrm{PSM}}=\int_{\mathcal{M}}\left[ A_I\wedge \dd X^I +\tfrac 12\Pi^{IJ}(X)A_I\wedge A_J\right]\label{IPSM}
\end{equation}
Note that Poisson sigma models may be formulated also without relying on the choice of a local coordinate system on $\mathcal{P}$ \cite{Bojowald:2004wu}

Through the identifications $ X^I=(X,X^\alpha)$ and $A_I=(\omega,e_\alpha)$ we see that the dilaton gravity action (\ref{D02}) is a particular case of Poisson sigma model action corresponding to
\begin{equation}
\Pi^{\alpha\beta}=-\mathcal{V}\epsilon^{\alpha\beta},\qquad
\Pi^{X\beta}=-\Pi^{\beta X}=X_\alpha\epsilon^{\alpha\beta}. \label{Pi}
\end{equation}
It is an easy exercise to check that the Poisson tensor (\ref{Pi}) satisfies the Jacobi identities (\ref{Jac}). 

Under a change of the target space coordinates $X^I\to X'^{I'}$ the fields $A$ and $\Pi$ change according to the rule 
\begin{equation}
A'_{I'}=A_I \frac{\partial{X^I}}{\partial{X'_{I'}}},\qquad
\Pi^{I'J'}(X')=\Pi^{IJ}(X(X'))\frac{\partial X'^{I'}}{\partial X^{I}}\, \frac{\partial X'^{J'}}{\partial X^{J}}.\label{prime}
\end{equation}

The PSM gauge transformations read
\begin{eqnarray}
&&\delta_\lambda X^I=\Pi^{IJ}\lambda_J\,,\label{gautr}\\
&&\delta_\lambda A_I=-\dd \lambda_I -(\partial_I\Pi^{JK})A_J\lambda_K,\nonumber
\end{eqnarray}
where $\lambda_I$ is a parameter. The relations between these transformations and gauge symmetries of dilaton gravities will be explained in the next section.

\section{Asymptotic equivalence}\label{sec:asy}
Given a Poisson manifold $\mathcal{P}$, symplectic leafs are defined as submanifolds in $\mathcal{P}$ to which the Hamiltonian vector fields $\Pi^{IJ}\partial_J F(X^K)$ are tangential at each point for any smooth function $F$. Symplectic leaves are even-dimensional. Since for Poisson sigma models associated with dilaton gravities $\mathrm{dim}\, \mathcal{P}=3$, the symlectic leaves may have dimension 2 or 0. 

Consider the function
\begin{equation}
\mathcal{C}=w(X)+\tfrac 12 X^\alpha X_\alpha e^{Q(X)} \label{C}
\end{equation}
with
\begin{equation}
Q\left(X\right)=\int^{X}\mathrm{d}y\;U\left(y\right),\qquad w\left(X\right)=\int^{X}\mathrm{d}y\;V\left(y\right)e^{Q\left(y\right)}.    \label{Qw}
\end{equation}
Since
\begin{equation}
\{ \mathcal{C},X^I\}=0,
\end{equation}
$\mathcal{C}$ is a Casimir function. It is constant on each symplectic leaf. Moreover, the PSM equations of motion yield $\dd \mathcal{C}(X,X^\alpha)=0$. Thus any classical solution of a PSM always stays within a single symplectic leaf.

Two-dimensional symplectic leaves\footnote{0-dimensional symplectic leaves are the points where $X^1=X^2=V(X)=0$. The classical solutions belonging to such leaves are the constant dilaton solutions. Their (somewhat trivial) holographic aspects have been discussed in \cite{Grumiller:2015vaa,Cvetic:2016eiv}.} are the surfaces $\mathcal{C}=\mathrm{const}$. Note, however, that the same value of $\mathcal{C}$ may correspond to several symplectic leaves. Locally on these surfaces one introduces the coordinates $(X,\theta)$, where
\begin{equation}
\theta=\mathrm{arctan}(X^2/X^1). \label{theta}
\end{equation}
It can be easily verified that
\begin{equation}
\{ X,\theta\}=1.\label{Xtheta}
\end{equation}
Thus, $(X,\theta)$ form a Darboux coordinate system on symplectic leaf. The coordinate systems consisting of Casimir functions and Darboux coordinates on symplectic leaves are called the Casimir-Darboux coordinate systems.

Let us consider a dilaton gravity model with an asymptotic region at $X\to\infty$. In this work, this means that for each value of $\mathcal{C}$ there is a critical value $X$, $X_{\mathrm{cr}}(\mathcal{C})$ such that at $X>X_{\mathrm{cr}}$ the equation (\ref{C}) has a unique positive solution for $X_\alpha X^\alpha$ smoothly depending on $\mathcal{C}$ and $X$. By the construction, the line $X^\alpha X_\alpha =0$ is excluded from the asymptotic region.
In this region, the coordinates $(X,\mathcal{C},\theta)$ form a regular coordinate system in the configuration space. Thus, by making the change of variables $(X,X^1,X^2)\to (X,\mathcal{C},\theta)$ we can establish a correspondence between asymptotic conditions in our model and in the PSM with the action
\begin{equation}
 I_{\mathrm{PSM}_0}=\int_{\mathcal{M}}\left[ A_C\wedge\dd\mathcal{C}+A_\theta\wedge\dd \theta +A_X\wedge\dd  X +A_X\wedge A_\theta \right]. \label{PSMprime} 
\end{equation}
Note, that this model does not correspond to any dilaton gravity.

In 2D dilaton gravities, the dilaton field itself can be used as a coordinate in classical solutions. Thus by defining as asymptotic region in terms of $X$ we also define an asymptotic region on $\mathcal{M}$.

Consider two dilaton gravity models with target space coordinates $(X,X^1,X^2)$ and $(Y,Y^1,Y^2)$ and with Poisson tensors $\Pi_{(X)}(X^I)$ and $\Pi_{(Y)}(Y^I)$, respectively. If both models have asymptotic regions at $X\to\infty$ and $Y\to\infty$, by the change of variables $(X,X^1,X^2)\to (X,\mathcal{C},\theta)\to (Y,Y^1,Y^2)$  one establishes an asymptotic correspondence between these models. This correspondence is valid in the intersection of asymptotic regions of the models. We shall call this intersection the asymptotic region until the end of this section.

By the construction, the Poisson tensors $\Pi_{(X)}(X^I)$ and $\Pi_{(Y)}(Y^I)$ in asymptotic regions and the one-forms $A^{(X)}$ and $A^{(Y)}$ are related through the equations (\ref{prime}). Now, we are going to establish relations between boundary actions, variational problems, and asymptotic symmetries.

To ensure consistency of the variational principle for a given set of asymptotic conditions, one had to add to $I_{\mathrm{PSM}}$ a boundary action $I_{\mathrm{bd}}$ defined at $X\to\infty$, so that the full action reads 
\begin{equation}
I=I_{\mathrm{PSM}}+I_{\mathrm{bd}}.\label{III}
\end{equation}
One may like to identify full actions in both models $I_{(X)}(X(Y),A^{(X)}(A^{Y},Y))=I_{(Y)}(Y,A^{(Y)})$ before and after the change of variables. This is not possible however since the transformation $X\to Y$ is defined in the asymptotic region only. Thus, we have to assume that there is some asymptotic region also in the base manifold $\mathcal{M}$ such that for $x$ belonging to this asymptotic region the fields $X^I(x)$ are in the asymptotic region of target space $\mathcal{P}$. This assumption will be a part of the asymptotic conditions. The principle describing the correspondence between classical actions is formulated as follows. Consider four sets of the fields, $(X,A^{(X)})$, $(\bar X,\bar A^{(X)})$, $(Y,A^{(Y)})$, $(\bar Y,\bar A^{(Y)})$, such that (i) the pair-wise differences $X-\bar X$, $A^{(X)}-\bar A^{(X)}$, $Y-\bar Y$, $A^{(Y)}-\bar A^{(Y)}$ have support inside the asymptotic region in $\mathcal{M}$, and (ii) the fields $(X,A^{(X)})$ (respectively, $(\bar X,\bar A^{(X)})$) are related to $(Y,A^{(Y)})$ (respectively, to $(\bar Y,\bar A^{(Y)})$) through in the diffeomorphsms (\ref{prime}) within the asymptotic region. Then
\begin{equation}
I_{(X)}(X,A^{(X)})-I_{(X)}(\bar X,\bar A^{(X)})=I_{(Y)}(Y,A^{(Y)})-I_{(Y)}(\bar Y,\bar A^{(Y)}). \label{IXIY}
\end{equation}
In other words, the variations of both actions should agree provided these variations are confined to asymptotic regions.

It can be checked by a direct computation that the bulk action $I_{\mathrm{PSM}}$ satisfies the condition (\ref{IXIY}). This allows au to identify the boundary actions as well
\begin{equation}
I_{\mathrm{bd}(X)}(X(Y),A^{(X)}(A^{Y},Y))=I_{\mathrm{bd}(Y)}(Y,A^{(Y)}).\label{IbdIbd}
\end{equation}
This boundary action is well defined since the asymptotic boundary clearly belongs to the asymptotic region. The relations (\ref{IXIY}) and (\ref{IbdIbd}) allow us to conclude that the equations in both theory agree in the asymptotic region. The asymptotic conditions in both models are also identified through the target space diffeomorphisms. If the variational problem for one model is consistent, the variational problem for the other is also consistent.

The asymptotic symmetries are the gauge symmetries of bulk theory which preserve asymptotic conditions but are no longer gauge symmetries of the full theory as they change the boundary action and correspond to non-zero but finite asymptotic charges. Thus, to understand the asymptotic symmetries we have to find out how the gauge symmetries change under target space diffeomorphisms. One may expect a tensorial behaviour  as
\begin{equation}
\delta_{\lambda} X^I=\frac{\partial X^I}{\partial X'^{J'}}\ \delta_{\lambda'}X'^{J'},\qquad \lambda'_{I'}=\frac{\partial X^J}{\partial X'^{I'}}\, \lambda_J.\label{deldel}
\end{equation}
(We have returned to simpler notations compatible with Eqs.\ (\ref{prime}) and (\ref{gautr}) above.) These rules are obviously consistent with transformation of $X$ and $\Pi$. However, the compatibility between gauge and diffeomorphism transformations of $A$ requires equations of motion. For us, this means that the asymptotic symmetries in the PSM models related by through asymptotic target space diffeomorphism coincide on shell only.

The necessity of going on shell is related to non-covariance of the variables used in PSMs. The field $X^I$ is a coordinate on $\mathcal{P}$ rather than a vector in the tangent space to $\mathcal{P}$. This is exactly the reason for non-covariance of gauge transformations with respect to the tangent space diffeomorphisms. The situation improves on shell, where the gauge transformations become covariant and close with respect to the commutator. This situation was discussed in detail in \cite{Bojowald:2004wu} where also a covariant extensions of PSMs were proposed. 

On shell, the PSM gauge transformations can be related to diffeomorphism of $\mathcal{M}$ with the parameters $\xi^\mu$ and Euclidean Lorentz transformations with the parameter $\sigma$ which constitute the symmetries in geometric formulation of 2D dilaton gravities \cite{Strobl:1994yk,Grumiller:2002nm}, $\lambda_I=\lambda_I(\xi)+\lambda_I(\sigma)$. 
\begin{equation}
\lambda_I(\xi)=-A_{\mu I}\xi^\mu,\qquad \lambda_X(\sigma)=\sigma,\qquad \lambda_\alpha(\sigma)=0.\label{lambdaxi}
\end{equation}
To restore $\xi$ it is enough to use the equation 
\begin{equation}\lambda_\alpha=-e_{\alpha\mu}\xi^\mu. \label{xie}
\end{equation} 
Since $\partial X/\partial X'^\alpha=0$ both sides of this equation transform identically and do not mix with the $\lambda_X$ component under the target space diffeomorphisms. Therefore, $\xi^\mu$ does not transform,
\begin{equation}
\xi^{'\mu}=\xi^\mu . \label{xixi}
\end{equation}  
This means in particular that in asymptotically equivalent models asymptotic Killing vectors coincide. This is a remarkably simple relation.

\section{An example: JT and asymptotically Rindler models}\label{sec:Ex}
In this section, we show how the known asymptotic conditions of JT gravity can be mapped to a new set asymptotic conditions of asymptotically Rindler models. 
\subsection{The JT model}\label{sec:JT}
The JT gravity corresponds to the dilaton potentials $V(X)=-X$ and $U(X)=0$. The Casimir function reads
\begin{equation}
\mathcal{C}=\tfrac 12 (-X^2+X_\alpha X^\alpha).\label{JTC}
\end{equation}
The two-dimensional symplectic leaves are paraboloids, except for $\mathcal{C}=0$, when they are two cones without the point $X=X^1=X^2=0$ (which is a zero-dimensional leaf). For a positive $\mathcal{C}$ there is single symplectic leaf, while for $\mathcal{C}<0$ there are two simplectic leaves. Obviously, for $X>X_{\mathrm{cr}}=\sqrt{-2\mathcal{C}}$ when $\mathcal{C}\leq 0$ and for all $X$ when $\mathcal{C}>0$, the equation (\ref{JTC}) has a unique positive smooth solution for $X_\alpha X^\alpha$. Thus, the JT model has an asymptotic region at $X\to\infty$. 

The asymptotic conditions for JT model in the first-order formulation have been constructed in a number of papers \cite{Grumiller:2013swa,Grumiller:2015vaa,Grumiller:2017qao}. We give a short overview here. The metric is taken in the Fefferman-Graham form
\begin{equation}
\dd s^2=\dd\rho^2+h^2(\tau,\rho)\dd\tau^2,\label{s2JT}
\end{equation}
and the zweiben reads
\begin{equation}
e_{\tau 1}=h,\qquad e_{\rho 2}=1.\label{eJT}
\end{equation}
The equations (\ref{s2JT}) and (\ref{eJT}) mean that the diffeomorphism and Lorentz gauges have been partially fixed. The asymptotic boundary corresponds to $\rho\to\infty$ with $\tau$ being a coordinate on this boundary. In the first-formalism, one can separate the equations of motions containing the derivatives w.r.t. $\rho$ and the ones containing $\partial_\tau$ only. The former are used to find a consistent set of asymptotic conditions. They yield
\begin{eqnarray}
&&\omega_\rho =0,\qquad \omega_\tau=\partial_\rho h,\\
&&X=x_R(\tau)e^\rho +x_L(\tau)e^{-\rho},\qquad
X^1=-x_R(\tau)e^\rho +x_L(\tau)e^{-\rho}, \label{asc1}\\
&&X^2=\ell(\tau),\qquad
h=e^\rho +\mathcal{L}(\tau)e^{-\rho}\label{asc2}
\end{eqnarray}
Here $x_R$, $x_L$, $\ell$, and $\mathcal{L}$ are arbitrary functions of $\tau$. They characterize the holographic theory. On shell, they obey the rest of the equations of motion containing $\tau$-derivatives (denoted by a dot over corresponding functions)
\begin{equation}
\dot{x}_R=\ell,\qquad \dot{x}_L=\ell\mathcal{L},\qquad \dot{\ell}=2(x_R\mathcal{L}+x_L)\label{taueomJT}
\end{equation}
The asymptotic conditions consist in requesting that for $\rho\to\infty$ the corrections to $X$, $X^1$, $X^2$ and $h$ are smaller than the smallest function of $\rho$ appearing on the right hand sides of the equations (\ref{asc1}) and (\ref{asc2}). The same apply to the formulas for transformation parameters given below. We do not write the correction terms explicitly in this subsection. As a part of the asymptotic conditions, we also have to request that the fields $X^I$ belong to the asymptotic region of target space for sufficiently large values of $\rho$. Thus, $x_R$ has to be separated from zero, i.e. $0<c\leq x_R(\tau)$ for all values of $\tau$ and some constant $c$. Note that the consistency of variational problem requires that the average value of $x_R^{-1}$ over the boundary does not fluctuate \cite{Grumiller:2017qao}. This also implies that $x_R^{-1}$ is regular.

These asymptotic conditions are invariant under the gauge transformations (\ref{gautr}) with
\begin{eqnarray}
&&\lambda_1=\varepsilon \, e^{\rho} -\bigl( \tfrac 12 \ddot{\varepsilon}-\mathcal{L}\varepsilon \bigr)e^{-\rho},\\
&&\lambda_2=-\dot{\varepsilon},\\
&&\lambda_X=\varepsilon\, e^{\rho} +\bigl( \tfrac 12 \ddot{\varepsilon}-\mathcal{L}\varepsilon \bigr)e^{-\rho},
\end{eqnarray}
where $\varepsilon$ is arbitrary function of $\tau$, cf \cite{Grumiller:2013swa}. The transformation law
\begin{equation}
\delta_\varepsilon\mathcal{L}=-2\mathcal{L}\dot{\varepsilon}-\dot{\mathcal{L}}\varepsilon +\tfrac 12 \dddot{\varepsilon} \label{JTVir}
\end{equation}
reveals Virasoro asymptotic symmetry with a non-vanishing central charge. Other asymptotic variables transform as
\begin{eqnarray}
&&\delta_\varepsilon x_R=x_R\dot{\varepsilon}-\ell \varepsilon,\\
&&\delta_\varepsilon x_L=-x_L\dot{\varepsilon}+\tfrac 12 \ell\ddot{\varepsilon}-\mathcal{L}\ell\varepsilon,\\
&&\delta_\varepsilon\ell=x_R(\ddot{\varepsilon}-2\mathcal{L}\varepsilon)-2x_L \varepsilon.
\end{eqnarray}

It is important to note that we have 3 equations of motion (\ref{taueomJT}) for 4 independent functions on the boundary. One of these functions, say $\mathcal{L}$, remains unrestricted even on shell. According to Eq.\ (\ref{JTVir}), this boundary degree of freedom may be described in terms of the orbits of Virasoro group. 

The asymptotic Killing vectors are the generators of diffeomorphisms of $\mathcal{M}$ which preserve asymptotic conditions for the metric. They read in our case
\begin{equation}
\xi^\rho=\dot{\varepsilon},\qquad \xi^\tau=-\varepsilon +\tfrac 12 \ddot{\varepsilon}\, e^{-2\rho}. \label{JTKilling}
\end{equation}

We conclude this subsection with a simple example of the behaviour of gauge symmetries under target space diffemorphisms. By using the rules (\ref{prime}), one gets for the connections in the model (\ref{PSMprime})
\begin{equation}
A_\theta=A_1\frac{\partial X^1}{\partial\theta}+A_2\frac{\partial X^2}{\partial\theta}=-e_1 X^2+e_2X^1, 
\end{equation}
which yields in the asymptotics
\begin{equation}
A_{\rho,\theta}\simeq -x_Re^\rho +x_Le^{-\rho},\qquad
A_{\tau,\theta}\simeq -\ell e^\rho -\ell \mathcal{L} e^{-\rho}
\end{equation}
Then,
\begin{equation}
\delta_\varepsilon A_{\rho,\theta}\simeq -(x_R\dot{\varepsilon}-\ell\varepsilon)e^{\rho}+\dots ,\qquad \delta_\varepsilon A_{\tau,\theta}\simeq (-x_R\ddot{\varepsilon}+2x_R\mathcal{L}\varepsilon+2x_L \varepsilon)e^\rho +\dots
\end{equation}
The first of the equations above is a gauge transformation with $\lambda_\theta=(x_R\dot{\varepsilon}-\ell\varepsilon)e^{\rho}+\dots$ However, the variation $\delta_\varepsilon A_{\rho,\theta}$ is reproduced only if one takes into account the equations of motion (\ref{taueomJT}).

We do not write explicitly the asymptotic conditions for other fields in the Casimir-Darboux coordinates $(X,\mathcal{C},\theta)$. Instead, in the next subsection we will directly transform the JT variables to asymptotically Rindler variable. This latter way appears to be shorter and simpler.

\subsection{Asymptotically Rindler models}\label{sec:Rin}

Among 2D dilaton gravities where is an interesting family with asymptotically Rindler solutions. The Euclidean version of this family corresponds to the potentials
\begin{equation}
U(Y)=-\,\frac{a}{Y},\qquad V(Y)=-\tfrac 12 Y^a, \label{UVR}
\end{equation}
where $a$ is a real parameter. Then
\begin{equation}
Q(Y)=-a\ln Y,\qquad w(Y)=-\tfrac 12 Y,\qquad \mathcal{C}=\tfrac 12 \bigl(-Y+Y^\alpha Y_\alpha Y^{-a}\bigr)\label{QwCR}
\end{equation}
The curvature scalar on the solutions is $R=-2a\mathcal{C}Y^{a-2}$. Thus, for $a<2$ the region $Y\to\infty$ is asymptotically flat. 

Our method allows to find asymptotic conditions with Virasoro symmetry in asymptotically Rindler models.
In the known cases \cite{Ecker:2021guy} these models in 2D have a twisted warped conformal asymptotic symmetry algebra\footnote{See also \cite{Afshar:2021qvi} where the BMS asymptotic symmetry in flat JT model was discussed in detail.  }. Being written through the Fourier modes of generators $T$ and $P$, this algebra reads
\begin{eqnarray}
&& i[T_n,T_m]=(n-m)T_{n+m},\nonumber\\
&& i[T_n,P_m]=-m P_{m+n} +i k (n^2-n)\delta_{m+n,0},\nonumber\\
&& i[P_n,P_m]=0.\label{warpedconformal}
\end{eqnarray}
The only central extension with a central charge $k$ appears in the $[T_n,P_m]$ commutator. This central extension cannot be moved to other commutators by means of a change of the basis. This algebra has a Schwarzian \cite{Afshar:2019tvp} which differs from that of Virasoro algebra and the corresponding Schwarzian action is a limiting case \cite{Afshar:2019axx} of complex SYK model \cite{Maldacena:2016hyu,Davison:2016ngz,Chaturvedi:2018uov,Gu:2019jub}. Thus, the known cases are very different from what we suggest here.

Since we consider non-integer values of $a$, negative values of $Y$ must be excluded from the beginning. For $Y>-2\mathcal{C}$, the quantity $Y^\alpha Y_\alpha$ is a smooth positive function of $Y$ and $\mathcal{C}$, so that the models under consideration have an asymptotic region at $Y\to\infty$.

We identify $X=Y$ and equate $\mathcal{C}$ in both models to obtain
\begin{equation}
X^\alpha X_\alpha=X^2-X+Y^\beta Y_\beta X^{-a},\qquad Y^\beta Y_\beta =X^a (X-X^2+X^\alpha X_\alpha).
\end{equation}
Equating the angles $\theta$ in both models yields
\begin{equation}
X^\alpha=Y^\alpha\sqrt{\frac{X^2-X+Y^\beta Y_\beta X^{-a}}{Y^\gamma Y_\gamma}},\qquad
Y^\alpha=X^\alpha\sqrt{\frac{X^a (X-X^2+X^\beta X_\beta)}{X^\gamma X_\gamma}}
\end{equation}
meaning the following asymptotic expansion for $Y^\alpha$
\begin{eqnarray}
Y^1 &=&	-x_{R}^{\frac{1}{2}\left(a+1\right)}e^{\frac{1}{2}\left(a+1\right)\rho}\left[1+ \frac{(\ell^2 - 4x_L x_R)}{2x_R} e^{-\rho}+ \mathcal O\left(e^{-2\rho}\right)\right],\\
Y^{2} &=& \ell x_{R}^{\frac{1}{2}\left(a-1\right)}e^{\frac{1}{2}\left(a-1\right)\rho}\left[1+ \frac{(\ell^2 - 4x_L x_R)}{2x_R} e^{-\rho}+ \mathcal O\left(e^{-2\rho}\right)\right].
\end{eqnarray}
The asymptotic conditions for connections $A$ in the Rindler models are the target space diffemorphisms of corresponding conditions in JT. After long but otherwise straightforward computations one obtains for the spin connection
\begin{eqnarray}
{\omega}_{\rho}	&=&	\ell\left[ \frac{1}{x_R}e^{-\rho}-\frac{\left(1+a\right) }{2 x^2_R}e^{-2\rho}+\mathcal{O}\left(e^{-3\rho}\right)\right],\\
{\omega}_{\tau}	&=&	\frac{\left(1+a\right)}{2x_R}+\frac{\left[(2+a)\ell^2-4(1+a)x_Lx_R-4\mathcal L x^2_R\right]}{2x^2_R}e^{-\rho} + \mathcal{O}\left(e^{-2\rho}\right).
\end{eqnarray}
The asymptotic expansions for zweibein components read
\begin{eqnarray}
{e}_{1\rho}	&=& \ell x_{R}^{-\frac{1}{2}\left(a+3\right)}e^{-\frac{1}{2}\left(a+3\right)\rho}\left[x_{R} e^\rho-\frac{1}{2}\left(2+\ell^{2}-4x_{L}x_{R}\right)+\mathcal{O}\left(e^{-\rho}\right)\right],\\
{e}_{1\tau}&=&x_{R}^{-\frac{1}{2}\left(a-1\right)}e^{-\frac{1}{2}\left(a-1\right)\rho}\left[\frac{1}{x_{R}}+\frac{\left(3\ell^{2}-4x_{L}x_{R}\right)}{2x_{R}^{2}}e^{-\rho}+\mathcal O(e^{-2\rho})\right],\\
{e}_{2\rho}	&=&x_{R}^{-\frac{1}{2}\left(a-1\right)}e^{-\frac{1}{2}\left(a-1\right)\rho}\left[1-\frac{\left(\ell^{2}-4x_{R}x_{L}\right)}{2x_{R}}e^{-\rho}+\mathcal O(e^{-2\rho})\right],\\
{e}_{2\tau}	&=& \ell x_{R}^{-\frac{1}{2}\left(a+3\right)}e^{-\frac{1}{2}\left(a+3\right)\rho}\left[x_{R}e^{2\rho}-\frac{\left(2+\ell^{2}-4x_{L}x_{R}\right)}{2}e^\rho + \mathcal{O}\left(1\right)\right].
\end{eqnarray}
In this subsection, the fields $\omega_\mu$ and $e_\mu$ always refer to asymptotically Rindler models. We do not introduce any special notation for them.

The parameters of PSM asymptotic symmetry transformations take the form
\begin{eqnarray}
{\lambda}_{Y} &=& \frac{1+a}{x_R}\varepsilon + \frac{\left(2+a\right)\ell^{2}\varepsilon-4\left(1+a\right)x_{L}x_{R}\varepsilon-4\mathcal{L}x_{R}^{2}\varepsilon-2\ell x_{R}\varepsilon+2x_{R}^{2}\ddot{\varepsilon}}{2x_{R}^{2}}e^{-\rho} + \mathcal{O}\left(e^{-2\rho}\right),\label{G07a}\\
{\lambda}_{1} &=& x_{R}^{-\frac{1}{2}\left(a-1\right)}e^{-\frac{1}{2}\left(a-1\right)\rho}\left[ \frac{1}{x_{R}}\varepsilon+\frac{3\ell^{2}\varepsilon-4x_{L}x_{R}\varepsilon-2\ell\dot{\varepsilon}}{2x_{R}^{2}}e^{-\rho}+\mathcal{O}\left(e^{-2\rho}\right) \right],\label{G07b}\\
{\lambda}_{2} &=& x_{R}^{-\frac{1}{2}\left(a-1\right)}e^{-\frac{1}{2}\left(a-1\right)\rho} \left[ \left(\frac{\ell}{x_{R}}\varepsilon-\dot{\varepsilon}\right)+\left(-\frac{\ell\left(2+\ell^{2}-4x_{L}x_{R}\right)\varepsilon}{2x_{R}^{2}}+\frac{\left(\ell^{2}-4x_{L}x_{R}\right)\dot{\varepsilon}}{2x_{R}}\right)e^{-\rho}+\mathcal{O}\left(e^{-2\rho}\right) \right].\label{G07c}
\end{eqnarray}

The first nontrivial check for these relations is that Eqs.\ (\ref{xie}) and (\ref{xixi}) indeed reproduce the asymptotic Killing vector (\ref{JTKilling}). One can also check that PSM gauge transformations with the parameters (\ref{G07a}) - (\ref{G07c}) match the world volume diffeomorphisms of $\mathcal{M}$,
\begin{equation}
\delta_\varepsilon\, g_{\mu\nu}=\mathfrak{L}_\xi g_{\mu\nu}, \label{Lieder}
\end{equation}
where $\mathfrak{L}$ denotes the Lie derivative. Both checks can be done with the help of Mathematica.

We like to use this opportunity to illustrate the differences between Killing vectors and asymptotic Killing vectors. The Virasoro asymptotic symmetry algebra has an $\mathfrak{sl}(2,\mathbb{R})$ subalgebra. In the JT model, the asymptotic Killing vectors corresponding to this subalgebra can be extended to global Killing vectors of the AdS$_2$ space of $\mathcal{C}=0$ solution. These Killing vectors leave the metric invariant, but they change the dilaton. Since the metric of asymptotically Rindler models depends on the JT metric and the dilaton $X$, this metric is no longer invariant under $\mathfrak{sl}(2,\mathbb{R})$ transformations. In other words, the asymptotic Virasoro symmetry algebra of course contains an $\mathfrak{sl}(2,\mathbb{R})$ subalgebra also in the asymptotically Rindler case, but this subalgebra does not correspond to any Killing vectors of the solutions. 

To obtain a suitable boundary action one has to take the boundary action for the first-order JT gravity in one of available forms \cite{Grumiller:2017qao,Gonzalez:2018enk} and transform it to new variables according to (\ref{IbdIbd}). Note that our bulk action differs from the one used in \cite{Grumiller:2017qao,Gonzalez:2018enk} by an integration by parts and thus by an additional boundary integral of $X^I A_I$, which also has to be transformed. The resulting expression is quite complicated and not very instructive, so that we do not write it explicitly. 

One cannot identify full actions (\ref{III}) in both models. Only the variations of bulk actions are equal, see (\ref{IXIY}). Let us \emph{assume}, however, that both actions coincide on shell,
\begin{equation}
I_{\mathrm{JT}}\vert_{\mathrm{on\ shel}}=I_{\mathrm{Rindler}}\vert_{\mathrm{on\ shel}}.\label{IIonshell}
\end{equation}
Practically, we need this equation to hold between the parts of full actions which depend on asymptotic variables. The on-shell action for JT model has been calculated in \cite{Grumiller:2017qao}:
\begin{equation}
I_{\mathrm{JT}}\vert_{\mathrm{on\ shel}}=\frac{1}{2}\int\mathrm{d}\tau\;\frac{1}{x_{R}}\left[\left(\dot{x}_{R}\right)^{2}-2\ddot{x}_{R}x_{R}+4x_{R}^{2}\mathcal{L}\right].\label{IJTonshell}
\end{equation}
As we have mentioned above, even on-shell one of the asymptotic variables remains free. Let this variable be $x_R$. Let $\mathcal{L}$ be a constant representative of a Virasoro orbit. After introducing a new boundary coordinate $v$ such that $\dot{v}=x_R^{-1}$ and some elementary calculations (see in \cite{Ecker:2021guy}), one obtains
\begin{equation}
I_{\mathrm{JT}}\vert_{\mathrm{on\ shel}}=\int \dd v \left[ 2 \mathcal{L}(\partial_v \tau)^2 -\mathrm{Sch}[\tau](v)\right],\qquad \mathrm{Sch}[\tau](v)\equiv \frac{\partial_v^3 \tau}{\partial_v\tau}-\frac 32\, \left( \frac{\partial_v^2\tau}{\partial_v\tau} \right)^2,\label{Sch}
\end{equation}
which is the Schwarzian action. Thus the conjecture (\ref{IIonshell}) leads to an asymptotic Schwarzian action for Rindler models and our asymptotic conditions.

The asymptotic correspondence works both ways. One may use the twisted warped conformal asymptotic conditions of Rindler models \cite{Ecker:2021guy} to define JT asymptotic conditions with the same symmetry. 

\section{Other models}\label{sec:oth}
In order to understand the limits of applicability of our method let us consider a family of dilaton potentials
\begin{equation}
U(Y)=-\frac aY,\qquad U(Y)=-\frac B2 \, Y^{a+b} \label{abfam}
\end{equation}
depending on two real parameters $a$ and $b$ and on a scale factor $B$. We refer to \cite{Grumiller:2002nm} for a detailed description of this family and further references. For this family, the Casimir function reads
\begin{equation}
\mathcal{C}= - \frac{B}{2(b+1)}\, Y^{b+1} +\frac 12 Y^\alpha Y_\alpha\, Y^{-a}. \label{abC}
\end{equation}
This equation can be used to express $Y_\alpha Y^\alpha$ through $\mathcal{C}$ and $Y$. A unique positive smooth solution exists for any $\mathcal{C}$ and $Y$ larger than some critical value if
\begin{equation}
b+1>0 \quad \mbox{and}\quad B>0.\label{abbB}
\end{equation}
Any model which satisfies these restrictions has an asymptotic region at $Y\to\infty$. This includes the JT gravity ($a=0$, $b=1$, $B=2$) and the asymptotically Rindler models ($b=0$, $B=1$) which we have considered above. Besides, all spherically reduced gravities and the Callan-Giddings-Harvey-Strominger \cite{Callan:1992rs} model also satisfy (\ref{abbB}). If the asymptotic conditions, the boundary action, and the asymptotic symmetry algebra are known for one of the models satisfying (\ref{abbB}), they can be immediately translated to all other models in this class.

We conclude, that the method proposed is indeed very general.

\section{Conclusions}\label{sec:con}
In this section, we briefly describe how our method can be improved and generalized.

The necessity to impose equations of motion in order to achieve equivalence of the asymptotic symmetry algebras poses a limitation to the applicability of our results. This limitation can be probably removed in a covariant extension of PSMs \cite{Bojowald:2004wu}. Reformulation of known holographic results including the consistent variational problem in a covariant way even for JT requires much extra work which, however, may be worth doing.

Instead of identifying the variables $(X,\mathcal{C},\theta)$ one can equate any other pairs of coordinate systems as long as they have identical Poisson brackets. In particular, the Casimir function in one model can be identified with some expression depending on the Casimir function in the other model. In some cases, this can bring at least technical advantages. Even more, instead of identifying asymptotic regions, one may identify near-horizon regions to study the relations between near-horizon symmetries.

The method which has been described above can be extended with very few modifications to  holographic correspondence in other 2D models admitting a PSM description as higher-spin models \cite{Alkalaev:2013fsa,Grumiller:2013swa,Gonzalez:2018enk,Alkalaev:2019xuv} and dilaton supergravities in 2D \cite{Astorino:2002bj,Forste:2017kwy,Forste:2017apw,Cardenas:2018krd,Mertens:2020pfe}. It is also interesting to see how the quantum group symmetries \cite{Fan:2021bwt} transform under the target space diffeomorphisms. The inclusion of matter fields in this approach is hardly possible since the matter couplings are not covariant with respect to target space diffeomorphisms. We do not know yet how the canonical boundary charges of two-dimensional gravities (see, e.g., \cite{Ruzziconi:2020wrb} for a recent analysis) transform under the target space diffeomorphisms.

\begin{acknowledgments}
We are grateful to Hamid Afshar and Daniel Grumiller for discussions and previous collaboration. This work was supported in parts by the S\~ao Paulo Research Foundation (FAPESP), project 2016/03319-6, and by the grant 305594/2019-2 of CNPq.
\end{acknowledgments}

\bibliography{review,additions}

\end{document}